# Deep learning-based groupwise registration for longitudinal MRI analysis in glioma

Claudia Chinea Hammecher[1,2], Karin van Garderen[1,1a,6], Marion Smits[1,1a,2,6], Pieter Wesseling[3,4,] Bart Westerman[5], Pim French[6], Mathilde Kouwenhoven[7], Roel Verhaak[8,9], Frans Vos[1,2], Esther Bron[1], Bo Li[1*]

[1]Dept. of Radiology & Nuclear Medicine, Erasmus MC, Rotterdam, the Netherlands
[1a]Medical Delta, Delft, the Netherlands
[2]Dept. of Imaging Physics, Delft University of Technology, the Netherlands
[3]Dept. of Pathology, Amsterdam UMC / VU Medical Center, the Netherlands
[4]Laboratory for Childhood Cancer Pathology, Princess Máxima Center for Pediatric Oncology, Utrecht, the Netherlands
[5]Dept. of Neurosurgery, Amsterdam UMC / VU Medical Center, Amsterdam, the Netherlands
[6]Dept. of Neurology, Erasmus MC, Rotterdam, the Netherlands
[7]Dept. of Neurology, Amsterdam University Medical Centers / VU Medical Center, the Netherlands
[8]The Jackson Laboratory for Genomic Medicine, Farmington, United States
[9]Dept. of Neurosurgery, Amsterdam University Medical Centers / VU Medical Center, the Netherlands
*mail.van.boli@gmail.com

## Synopsis

Glioma growth may be quantified with longitudinal image registration. However, the large mass-effects and tissue changes across images pose an added challenge. Here, we propose a longitudinal, learning-based, and groupwise registration method for the accurate and unbiased registration of glioma MRI. We evaluate on a dataset from the Glioma Longitudinal AnalySiS consortium and compare it to classical registration methods. We achieve comparable Dice coefficients, with more detailed registrations, while significantly reducing the runtime to under a minute. The proposed methods may serve as an alternative to classical toolboxes, to provide further insight into glioma growth.

## Introduction

Glioma progression is monitored by routine MR scanning, enabling that tumor growth can be evaluated with respect to earlier time-points. This growth may present both as a mass effect and as an extension of abnormalities into previously healthy tissue. To accurately assess tumor growth and tumor-induced deformations, longitudinal intrasubject image registration is often used. However, such registration in cases with large deformations and tissue change is highly challenging.

Longitudinal image registration may benefit from groupwise strategies in which multiple images are concurrently aligned. This avoids introducing bias towards an a priori selected reference image. However, existing learning-based methods for image registration mostly concern pair-wise approaches[1]. Moreover, the few proposed learning-based methods for groupwise registration are designed for analysis of images without pathologies, and are prone to fail registering glioma images. To bridge this gap, we present a learning-based method for non-linear registration of longitudinal glioma images.

## Methods

We used T2-weighted FLAIR MRI scans of 61 participants from the multi-center GLASS-NL study[2]. Participants were initially diagnosed with lower-grade (grade 2 or 3) IDH-mutant astrocytoma and underwent multiple surgical resections. Images were affinely aligned to the ICBM 2009a nonlinear asymmetric atlas[3], skull-stripped and intensity normalized. We obtained tumor[4] and normal-appearing tissue segmentations[5]. For each subject, we grouped available scans before or after a surgical resection into all possible permutations of three time-points. The data was split into 46:15 patients (3349:90 permutations) for training and testing.

We expanded an existing learning-based registration approach[1] to take tumor presence and growth into account. During training, the method estimates the diffeomorphic deformations to the permutation's mean-space, maximizes the local cross-correlation across the warped images, and encourages a smooth and continuous deformation ($T_i$). To be robust against possible intensity alterations in the tumor region, a loss-function masking strategy was implemented to compute the loss value only in the normal-appearing region across the three time-points (H, Figure 2):

$$\bar{I} = \frac{1}{n}\sum_{i}^{n} I_i \circ T_i,$$
$$Loss = -\frac{1}{n}\sum_{i}^{n} L_{sim}(I_i \circ T_i, \bar{I}; H) + \lambda \cdot \frac{1}{n}\sum_{i}^{n} L_{reg}(T_i).$$

In addition, to register large local mass-effects caused by gliomas, we estimated the deformation at two resolutions, to firstly register the general structures in down-sampled images, and secondly refine the residual deformations at full resolution (Figure 1). We evaluated the proposed method against state-of-the-art classical groupwise registration methods: Elastix[6], NiftyReg[7], and ANTs[8]. These were run with default parameters, providing normal tissue masks as input when this option was available (i.e., Elastix and NiftyReg). The similarity across the warped images was assessed by the

Dice coefficients, and the average structural similarity index measure (SSIM) between warped image and the average image. Also, the centrality was evaluated by the average norm of the three resulting deformations. What is more, the smoothness of the deformations was measured by the number of foldings (negative values) in the Jacobian maps and their average standard deviation[6]. All metrics were computed in the normal-appearing tissue, and statistical significance with Wilcoxon signed-ranks over 15 independent subjects (p=0.05).

## Results

Figure 3 presents the average Dice and SSIM scores of all test permutations by the initial affine registration, the classical methods, and the proposed framework. Our single-stage method ('mask only') performed comparably to the classical methods in terms of Dice coefficient. The average SSIM obtained by our method was higher than for these classical methods, except Elastix. On the other hand, our multi-stage implementation ('mask+multi-stage') both Dice and SSIM coefficients with respect to the single-stage.

Elastix presents the best centrality, followed by our multi-resolution strategy (Table 1). The proposed strategies show significant improvement in smoothness and have inference runtimes of under a minute, significantly faster than the classical approaches. In a qualitative example (Fig. 4), the stronger deformations of Elastix lead to more overlap of the tumor across images, but with non-anatomically plausible deformations near the tumor edge. The proposed methods accurately align the normal-appearing tissue, but did not align the resection volume.

## Discussion

The proposed method is able to register glioma images despite the presence of non-correspondences across the time-points by focusing on the normal-appearing tissue similarity. The obtained GM and WM Dice coefficients are comparable to those of state-of-the-art toolboxes, but with higher SSIM values, suggesting that the registrations are more detailed. Elastix and NiftyReg show larger tumor Dice but stronger deformations, which could indicate anatomically implausible registration of non-correspondences. Qualitatively, our method shows stronger misalignment of the resection volume. This could indicate that changes in such volume are identified as non-correspondences instead of mass-effect.

Our method also achieves smoother deformations with the least foldings. An important advantage of our network approach is that new images can be registered in seconds, which is much faster than the classical methods (e.g., 28 hours by ANTs). We showed that the multi-stage strategy combined with the tumor masks yields higher registration accuracy than without this strategy, as this allows large, smooth deformations while avoiding local minima. However, for the cases with extremely large mass-effect, further refinement of the method could be considered.

## Conclusion

The proposed deep learning-based unbiased group-wise registration method can serve as an alternative to existing classical toolboxes for the analysis of glioma growth in longitudinal MRI.

## Acknowledgements

We would like to thank the members of the GLASS consortium, for the datasets and clinical insight. Also the members of the Biomedical Imaging Group Rotterdam for insightful dicussions. And in particular Dr. Frans Vos, Dr. Esther Bron, Dr. Bo Li and PhD. Karin van Garderen for their supervision and support during this research.

# Figures

**Table 1.** Centrality and smoothness of the estimated deformations, and the required runtime in minutes using the methods Elastix, NiftyReg, ANTs, and the proposed framework with and without multi-resolution implementation. Results are computed within the normal tissue and averaged over the test set. * indicates metrics that statistically significantly better than the classical methods, ** indicates those better than all the methods.

|  | ↓Centrality | ↓ Smoothness | | ↓ Run time [min] | Invertible |
|---|---|---|---|---|---|
|  |  | $|J_T| \leq 0$ [%] | SD $|J_T|$ |  |  |
| Elastix | 1.008e-14** | 6.665e-02 | 0.125 | 22.35 | x |
| NiftyReg | 4.828e-02 | 1.175e-02 | 0.159 | 32.58 | x |
| ANTs | 6.725e-02 | 3.161e-03 | 0.106 | 1647 | √ |
| Proposed: one resolution | 1.033e-03 | 0.000e+00* | 0.092 | < 1* | √ |
| Proposed: multi-resolution | 1.684e-03 | 5.999e-07 | 0.085** | < 1* | √ |

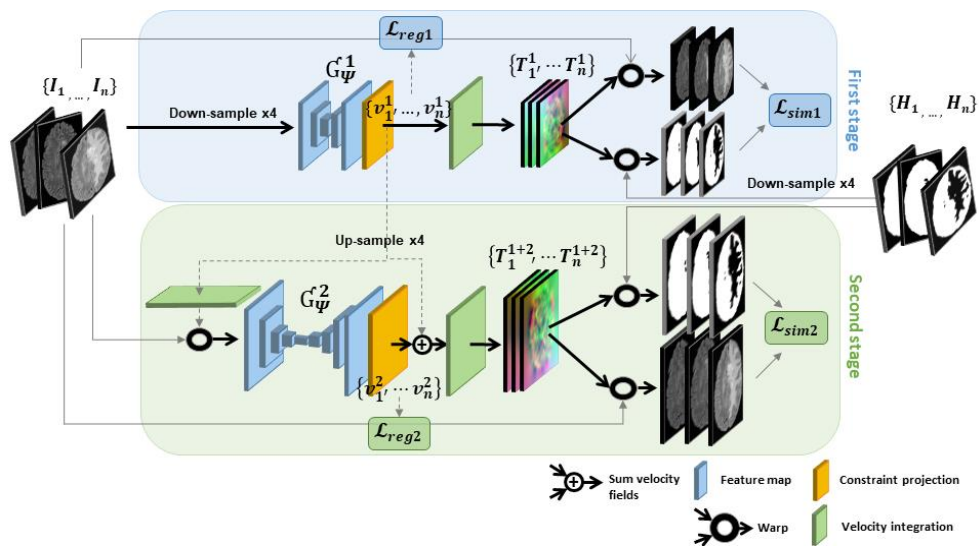

**Figure 1.** Schematic representation of the proposed method. The first stage is trained on the down-sampled FLAIR images $I_n$. After training, the velocity fields $v_n^1$ are up-sampled to warp $I_n$ and serve as an input to the second stage. The residual deformation fields are obtained from $v_n^1 + v_n^2$ and applied to $I_n$ to reduce interpolation error. In the second stage, the training parameters $G_\Psi^{\prime 1}$ are fixed. In both stages, the normal-appearing masks $H_n$ are included in the loss function.

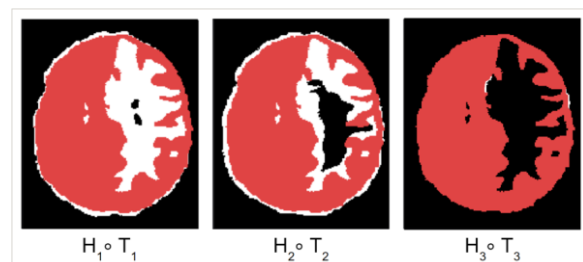

**Figure 2.** In white the normal-appearing region at three time-points (H1, H2, and H3), and in red their common region (H).

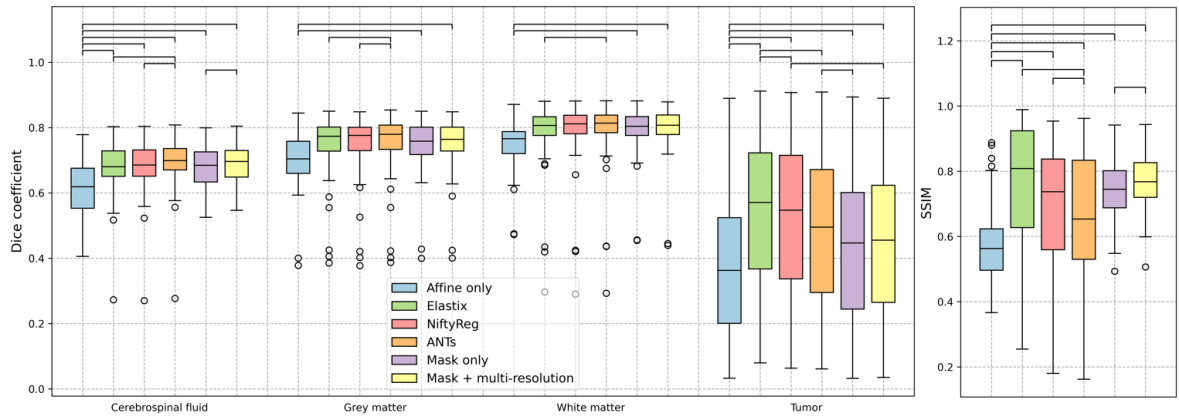

**Figure 3.** Registration accuracy for affine alignment only, Elastix, NiftyReg, ANTs, and our proposed method with and without multi-resolution implementation. Left: the boxplots of Dice coefficients for CSF, grey matter (GM), white matter (WM), and tumor region are shown. Right: the boxplots for average SSIM in normal-appearing tissue.

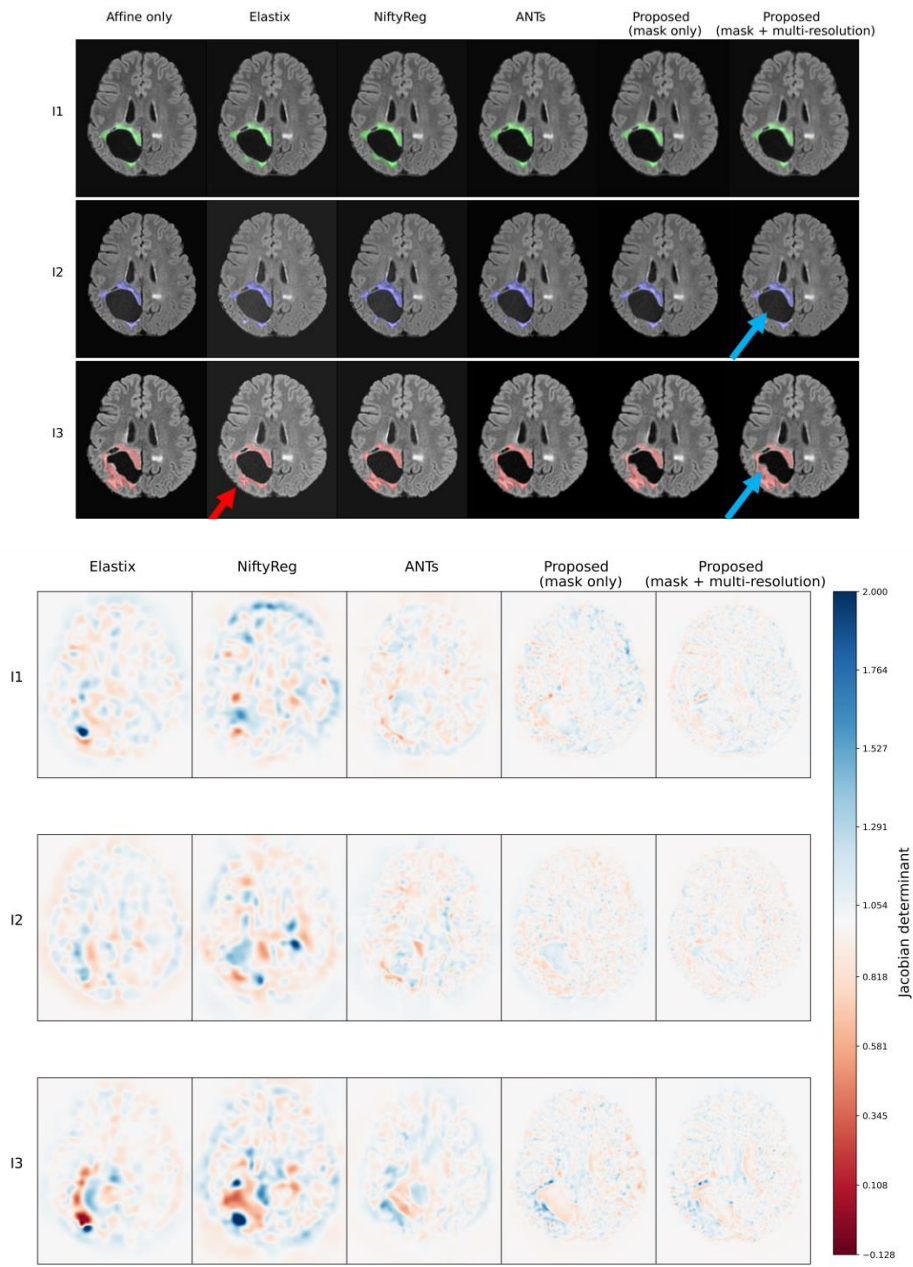

**Figure 4.** Results of one longitudinal permutation with images I1, I2, and I3 taken 3, 16, and 36 months after surgery. Overlaid the warped tumor segmentations. Red arrow: excessive compression of tumor. Blue arrows: resection cavity not aligned.